\documentstyle[aps,prl,preprint,psfig]{revtex}
\begin{document}
\draft          
\title{Branching Processes and Evolution at the Ends of a Food Chain}
\author{G. Caldarelli, C. Tebaldi}
\address{S.I.S.S.A./I.S.A.S., V. Beirut 2-4, I-34013
 Grignano di Trieste (TS), Italy}
\address{I.N.F.M., Istituto Nazionale di Fisica
della Materia, Sezione di Trieste}
\author{A. L. Stella}
\address{INFM--Dipartimento di Fisica e Sezione INFN,
Universit\`a di Padova,}
\address{I-35131 Padova, Italy}
\maketitle
\begin{abstract}       
In a critically self--organized model of punctuated
equilibrium, boundaries determine peculiar scaling
of the size distribution of evolutionary avalanches.
This is derived by an inhomogeneous generalization of standard
branching processes, extending previous mean field descriptions and 
yielding $\nu=1/2$ together with $\tau'=7/4$, 
as distribution exponent of avalanches starting from species at the ends of
a food chain. 
For the nearest neighbor chain one obtains numerically  
$\tau'=1.25 \pm 0.01$, and $\tau'_{first}=1.35 \pm 0.01$ for the 
first return times of activity, again distinct from bulk exponents.
\end{abstract}
Branching processes (BP) occur in many fields of physics and biology,
ranging from nuclear reactors, to polymers and population dynamics \cite{HAR}.
Within the context of self--organized criticality (SOC), introduced by
Bak, Tang and Wiesenfeld \cite{BTW1}, BP, or correlated versions of them,  
are expected to underlie the physics of many models, describing 
sandpiles \cite{BTW1,ALS,PVZ}, earthquakes \cite{OLA}, river 
networks \cite{SCH} or species mutations \cite{BS1}. 
By evolving long enough, these models self organize in stationary 
critical states with long--range
correlations in space and time, and with avalanches of activity occurring at
all scales. Avalanches are often believed to be described in terms of critical
BP in the mean field (MF) limit. In the present Letter we introduce and solve 
an inhomogeneus generalization of the standard BP. This allows us to
determine peculiar scaling properties of BP at boundaries. In a unifying
perspective, such properties provide a substantial extension of previous MF
descriptions of SOC models.

Bak and Sneppen (BS) \cite{BS1,BS2} introduced a SOC model describing an 
ecology of
interacting species evolving by mutation and selection. This model provides an
illustration of the mechanisms determining intermittency
(punctuated equilibrium \cite{GOULD}) and scaling \cite{RAUP} in 
the evolutionary activity.
Below we show that such intermittency and scaling have a richer structure than
appreciated so far. Indeed, at the level of universal properties, it is 
possible to draw a clear cut 
distinction between evolutionary activities occurring in the ``bulk"
and at the ``boundary" of an ecology. Bulk and boundary refer to different
locations of a given species within the network of interactions with
other species conditioning its evolution.

In a coarse grained, simplified description, BS associate to the 
$i$-th species of an ecology a single fitness parameter, $x_i$, ($0<x_i<1$).
$x_i$ represents the ability of species $i$ to survive:
the higher $x_i$, the higher the barrier to overcome in order to switch a
mutation in the species.
A genetic mutation changes the barrier of the species and modifies also the
barriers of the other species interacting directly  with it.
This interaction should represent the fact that two species, e.g., take
part in the same food chain. Sites of a lattice can be used to represent
the species: in this case nearest neighbor (n.n.) species can be assumed as
directly related biologically, and thus interacting.

The dynamical evolution rules are as follows. 
Starting from an initial fitness landscape, the $i$
with lowest $x$, $i_{min}$, is selected to undergo a mutation and its 
fitness $x_{i_{min}}$ is modified into a new one, chosen at random. 
Due to the interaction, also some neighbors of 
$i_{min}$ get modified $x$'s, as effect of the previous mutation. 
For a linear chain with n.n. interactions this implies that
$x_{i_{min}-1}$ and $x_{i_{min}+1}$ are replaced by new randomly
chosen $x$'s. In a standard MF description, on the other hand,
the notion of position is completely lost and one can, e.g., choose to replace
the fitnesses of a certain number, $K-1$, of other species
selected at random, besides $i_{min}$. This
random neighbor (r.n.) model is the
only one for which a MF treatment of avalanches could be set up so far  
\cite{BS2,DEB}. However, the lack of any meaning for distance in this MF 
is a quite strong limitation, to the extent that the very
notion of SOC can be legitimately questioned \cite{BJW}.

Avalanches correspond to sequences of mutations in which the minimal
$x$ species is always found among those resulting from genetic changes in
previous stages, starting from a given ancestor mutation with 
$x_{i_{min}}=\lambda$. 
In the system the minimal $x$ value does not exceed $\lambda$ for the
whole duration of the avalanche. 
The probability $P$ that an avalanche involves $s$ mutations is expected to
vary asymptotically as
$P(s) \propto s^{-\tau}$
in the SOC state, in which for all avalanches
$\lambda$ attains the value $x_c$, the sharp threshold of the stationary
$x$-distribution \cite{BS2}.

So far, in models like the n.n. chain, $\tau$ and similar exponents have 
always been discussed as bulk quantities \cite{BJW,PAC}, i.e. considering
statistics of avalanches starting everywhere within large, periodic systems.
Compared to those in the bulk, a species at one end
of an open chain (e.g., main predator, or basic
level of microscopic life) has less species directly
or indirectly connected to it. The paths through
which dynamical correlations can propagate starting from an initial
mutation on the boundary are also reduced. So, 
e.g., in semi-infinite geometry,
boundary avalanches could be characterized by peculiar exponents,
different from the bulk ones.
Demonstrating boundary scaling in models like the BS one is a
challenge, especially at the analytical level. Indeed, in the
context of SOC with extremal dynamics, exact results are essentially limited to
the above mentioned MF treatment \cite{DEB,BP}. Consideration 
of boundary effects
or other inhomogeneities clearly requires a meaningful notion of distance.
We achieve this within a novel MF description of the BS
model with n.n. interactions, generalizing the
standard BP studied in probability theory \cite{HAR}.

The main scaling result for the random neighbor MF model 
is $\tau=3/2$ \cite{BS2,DEB}.
This $\tau$ is consistent with MF BS avalanche dynamics being  
equivalent to a BP.
An avalanche can be identified with a tree, where nodes represent
species mutating within the avalanche. From each node as many branches depart,
as there are species undergoing genetic change directly due to a
mutation taking place at that node. The same species can act as node
more than once within an avalanche.
The complex structure of correlations of
the BS model is simplified in MF by assuming that, at each node, well defined,
independent probabilities exist for all branchings compatible with
the dynamical rules.
Avalanches are generation trees, whose distribution in number of generating
individuals, $s$, is given by $P$.
The existing MF approach clearly can not
address exponents for diverging lengths, as defined, e.g., in a
Landau approach to standard criticality.
We introduce a characteristic length within MF
through boundaries breaking translation invariance and
leading to a position dependence of the BP description.
Standard BP theory deals with the discrete transform $ \tilde{P}(z) =
\sum_{s=1}^{\infty} P(s)z^s $, on which the scaling of $P(s)$
produces singular behaviour of the form
\begin{equation}
\tilde{P}(z) = 1 - c (1-z)^{\tau -1} + l.s.t.
\label{eq:2}
\end{equation}
for $z \rightarrow 1^-$. In eq.(\ref{eq:2}) $c$ is a suitable 
positive constant and
the last term on the r.h.s. indicates regular or subleading singular terms.
Without making reference to relative locations of the species along the chain,
the standard BP assumes that well defined probabilities,
$p_i$ ($i=0,1,2,...K$), apply to the events in which a given species
undergoing mutation triggers subsequent genetic changes, in the same
avalanche, in $i$ species, possibly including itself.
Independence of branchings leads to validity of Watson's functional equation
\cite{HAR}
\begin{equation}
\tilde{P}(z) = zG(\tilde{P}(z))
\label{eq:3}
\end{equation}
with $G(y) = p_0 + p_1y + p_2y^2 + ...+ p_Ky^K$. Eq.(\ref{eq:3})
imposes a constraint on the $p_i$'s consistent
with a singularity of the form (\ref{eq:2}). Such constraint reads
$G^{\prime}(1)=1$ and automatically fixes $c=\sqrt{2/G''(1)}$ and $\tau=3/2$ 
as the only compatible exponent \cite{PVZ2}.
This result for $\tau$ is largely universal with
respect to different choices of the parameters $p_i$ and only relies
on the analyticity of $G$. A natural choice is 
$p_i=\left(  
\stackrel {\scriptstyle K} {_{_{\scriptstyle i} }} 
\right) 
x_c^i (1-x_c)^{K-i}$. 
In the r.n. model $x_c=1/K$ \cite{BS2},
implying satisfaction of $G'(1)=1$. Replacing $x_c$ by $\lambda<x_c$
would amount to consider off--critical avalanches ($G'(1)<1$), with
$x_c-\lambda$ playing the role of temperature-like field. 
Let us consider now a semi-infinite sequence
of species on a chain. To each species is associated an integer coordinate
$j=0,1,2,....$. In a n.n. model the presence of the boundary requires
to allow for a $j$--dependence of the avalanche size distribution:
thus, $P_j(s)$ or $\tilde{P}_j(z)$ will describe avalanches starting at
site $j$ along the chain. This situation 
can still be analysed within what we call here inhomogeneous BP.
Since, as a consequence of a mutation at $j \geq 1$, at most $3$ species
can be further involved in the avalanche ($K=3$ for the n.n. case), 
probabilities
$p_0,p_1,p_2$, and $p_3$ will describe the possible outcomes of such a
mutation. For convenience, and consistently with the
above expressions of the $p_i$'s in terms of $x_c$,
one can further assume that with probability
$p_0$, no further mutation takes place in the avalanche; with probabilities
$p_1/3$ and $p_2/3$ the avalanche propagates, respectively, in any one
and any two of the species in the set $\{j-1,j,j+1\}$; finally, $p_3$ is the
probability that the avalanche involves all three species.
In the MF spirit it is also sensible to assume
$j$-independence for the $p_i$'s as long as $j\geq 1$.
Of course, there should be different probabilities $p'_i$ for $j=0$, where the
boundary imposes $p'_3=0$. A possible choice made below is to assign
$p'_0=p_0 +\frac 1 3 p_1, p'_1=\frac 2 3 p_1 + \frac 2 3 p_2$ and 
$p'_2=\frac 1 3 p_2 +p_3$ at $j=0$, again
implying equivalence of $j=0$ and $j=1$ with respect to single branch
outcomes.

With the above positions, eq.(2) is replaced by a full hierarchy
of equations:
\begin{eqnarray}
\widetilde{P}_0(z)&=&z\left(p_0^{\prime }+\frac {p_1^{\prime }} 2 \left( 
\widetilde{P}_0(z)+\widetilde{P}_1(z)\right)+\right. \nonumber \\
& &\left. p_2^{\prime }\left(\widetilde{P}_0(z) \widetilde{P}_1(z)\right)
\right) \nonumber \\ 
\widetilde{P}_j(z)&=&z\left(p_0+\frac{p_1}3\left( 
 \widetilde{P}_{j-1}(z)+ \widetilde{P}_j(z)+ \right. \right. \nonumber \\ 
& &\left. \left.\widetilde{P}_{j+1}(z)\right)+ \frac{p_2}3 
\left(\widetilde{P}_{j-1}(z)\widetilde{P} _{j+1}(z)+\right. \right. \nonumber \\
& &\left. \left. \widetilde{P}_j(z)\widetilde{P} _{j+1}(z) +
\widetilde{P}_{j-1}(z)\widetilde{P}_j(z)\right)+\right. \nonumber \\
& &\left.p_3\left(\widetilde{P}_{j-1}(z)\widetilde{P}_j(z)
\widetilde{P}_{j+1}(z)\right)\right); \mbox{$j \geq 1$.}\label{eq:4}
\end{eqnarray}

$\widetilde{P}_j(z)$ should converge to the bulk solution of
eq.(\ref{eq:3}), for $j$ approaching infinity.
Thus, it is advantageous to adopt the following ansatz:
\begin{equation}
\tilde P_j(z)=\tilde P(z)+\Delta (z)e^{-q(z)j}+l.s.t. 
\label{eq:5}
\end{equation}
where $q$ is an inverse length and $\tilde{P}$ is the solution of 
eq.(\ref{eq:3}). As shown below, the assumed $j$-independence of
$\Delta$
and $q$ is consistent, as corrections to it would only involve subleading
singular terms for $z \rightarrow 1^-$.
By substituting eq.(\ref{eq:5}) into eqs.(\ref{eq:4}) one can deduce 
singular behaviours of $\tilde{P}_0$ and $q$. 
For $z \rightarrow 1^-$, we expect $\Delta(z) \sim (1-z)
^\alpha$ and $q(z) \sim (1-z)^\beta$, with $\alpha$ and $\beta$ suitable
exponents. After substitution in eqs.(\ref{eq:4}) for $j\geq1$ one gets 
\begin{eqnarray}
1&=&\frac z3(1+2\cosh {q(z)})\left( G^{\prime }\left( \tilde{P}\left(
z\right) \right) \right.+ \nonumber \\
 & & \left. \frac {\Delta(z)} 2 G^{\prime \prime }\left( \tilde{P} 
\left( z\right) \right) \right) +l.s.t. 
\label{eq:6}
\end{eqnarray}
Taking into account that $\tilde{P}$ has the form (\ref{eq:2}) with $\tau
= 3/2$, the leading singular terms in eq.(\ref{eq:6}) give:
\begin{equation}
\frac{q(z)^2}{3G^{\prime \prime }\left( 1\right)}+\frac 12\Delta
(z)= a(1-z)^{1/2}+l.s.t.; 
\label{eq:7}
\end{equation}
where $a=c$ of eq.(\ref{eq:2}).
The same kind of substitution in the first of eqs.(\ref{eq:4}) leads to
\begin{equation}
\Delta(z) = a(1-z)^{1/2}-b\Delta(z)q(z)+l.s.t.
\label{eq:9}
\end{equation}
with $b=1$.
Eq.(\ref{eq:7}) and eq.(\ref{eq:9}) determine both $\alpha$
and $\beta$ above. In particular $\tilde{P}_0(z)$ takes the form
\begin{equation}
\tilde{P}_0(z)\!=\!\tilde{P}(z)\!+\!\Delta(z)\!+l.s.t.\!=\!1\!-\!b(1-z)^{3/4}
\!+\!l.s.t.
\label{eq:10}
\end{equation}
In general $b=\frac {p'_1/2+p'_2} {2(p'_1/2+p'_2)-1}$
and the results (\ref{eq:7}) and (\ref{eq:10}) 
make sense for $\sum i p'_i<1$.
This condition is satisfied by our choice of $p'_i$'s, which further
acquire the form $p'_i=
\left(
\stackrel {\scriptstyle K-1} {_{_{\scriptstyle i} }} 
\right)
x_c^i(1-x_c)^{K-1-i}$, if the
$p_i$'s are expressed in terms of $x_c$ as discussed above. Thus, the
threshold $x_c$ for the distribution of $x$ values at the borders
in the stationary state should be the same as in the bulk.
According to eqs.(\ref{eq:2}) and (\ref{eq:10}) 
\begin{equation}
{P}_0(s) \sim s^{-7/4}
\label{eq:11}
\end{equation}  
Thus, in our MF description the BS SOC state is
characterized by a boundary scaling with an exponent $\tau'= 7/4$ 
different from the bulk one.
Boundary avalanches of course suffer more rapid extinction and their
distribution decreases faster for large $s$.
It is interesting to note that, by exploiting analogies with 
magnetic systems, $\tau' = 7/4$ has been predicted recently
within a MF approach to border avalanches in Abelian Sandpile Models (ASM) 
with Dirichelet boundary conditions \cite{STC}.
This lends further support 
to the idea that in ASM a BP
description underlies the statistics of avalanches in the MF limit, for
which also $\tau=3/2$ is expected \cite{MF}. By a numerical
approach one can also identify $\tau'\simeq 7/4$ for MF avalanches 
of the earthquake model of ref. \cite{OLA} \cite{LS}, confirming an underlying 
BP also in this case.
A further consequence of eqs.(\ref{eq:7}), (\ref{eq:9}) is the singularity:
\begin{equation}
q(z) \sim (1-z)^{1/4}.
\label{eq:12}
\end{equation}
Thus, the penetration length of the border disturbance, $q^{-1}$, diverges
for $z \rightarrow 1^-$. In MF treatments of inhomogeneous
equilibrium models, quantities like $q^{-1}$ show the same divergence with
temperature as classical correlation lengths. By interpreting $z$ as a standard
fugacity for a polymer, one deduces from eq.(\ref{eq:12}) a correlation
length exponent $\nu=1/4$. This is indeed the classical
$\nu$ of branched polymers \cite{HARL}.
Of course the definition of $\nu$ for a SOC system requires 
one to identify physically meaningful parameters describing
the approach or the departure from criticality. For BS avalanches such a
parameter is the temperature-like deviation $x_c-\lambda$.
By introducing $\lambda$-dependent $p_i$'s and $p'_i$'s
in our equations, the result (\ref{eq:12}) can be converted into
$q(\lambda,z=1) \sim (x_c-\lambda)^{1/2}$, which implies
$\nu=1/2$. Remarkably enough, this is the classical $\nu$
exponent expected for ASM \cite{MF}.
This and the above mentioned coincidence of $\tau'$ strongly support the
idea that BP fully underlie also the MF description of ASM avalanches.

In order to identify boundary scaling beyond MF, we performed systematic 
simulations with open, n.n. BS chains of different lengths ($N\leq 10^3$). 
First we verified that the distribution of boundary $x$'s in the stationary 
state is essentially unaltered with respect to that of the periodic, bulk case,
and displays the same sharp threshold at $x_c=0.665 \pm 0.015$ \cite{BS1}.
This coincidence is fully consistent with our choices of the $p'_i$'s
in the MF approach. 
By selectively sampling avalanches starting near the boundaries or in
the interior of the chains, we extrapolate $\tau'=1.25 \pm 0.01$ (see FIG. 1).
This value is clearly different from the bulk one $\tau \simeq 1.08$ 
\cite{BJW}.
So, also in the n.n.
model boundary avalanches have a probability decaying more
rapidly at large $s$, than in bulk. A further characterization of boundary 
scaling is given by the distribution of first return times of
activity ($x$ taking the minimum value) at the same boundary site. These times
are distributed as $t^{-\tau'_{first}}$, with $\tau'_{first}=1.35 \pm
0.01$, different from the bulk value $\tau_{first}=1.58$ \cite{PAC}. 
By recording the times of all subsequent returns of activity one can also 
obtain a distribution
$\propto t^{-\tau'_{all}}$, with $\tau'_{all}=0.65 \pm 0.01$, 
again distinct from 
$\tau_{all}=0.42$ in bulk \cite{PAC}. Such boundary exponents are consistent 
with a scaling relation $\tau'_{first}+\tau'_{all}=2$, already satisfied in 
the bulk \cite{PAC}. 
Since the validity of such relation should not depend on the position
considered along the chain, 
the above consistency is further indication of the good quality of our
determinations.
Data concerning these exponents are  shown in FIG. 2.

We conclude that at the boundaries activity has a 
different pattern of intermittency.
First returns are shifted towards longer time scales.
On the other hand, once the boundary has been reached, activity remains more easily 
trapped there, giving rise to concentrated sequences of returns.
In applications of the BS model, the choice of a more or less regular
network of interactions remains to some extent arbitrary, and should not
matter for universal properties, unless the long-range limit of a r.n.
model is assumed. However, the distinction elucidated above between bulk
and boundary species, appears to have important consequences, affecting 
the universal scaling features of evolution. Thus, boundary scaling 
offers additional, deeper
insight into the properties of biological models and widens
the context of their possible comparison with paleontological data. 
 Summarizing, we showed here that within the framework of punctuated
equilibrium there exists a well defined boundary scaling in addition
to the bulk one. At the MF level this scaling can be analysed exactly
within a generalization of BP theory, which considerably extends
previous classical descriptions of the BS model, and directly focuses on 
its relation with other models. In particular also ASM with Dirichlet 
boundary conditions fall fully in the MF universality class of our BP. 
 Also in the n.n. case our results 
show the existence of new scalings which make the notion of species at the 
ends of a chain meaningful in a universal sense.

\begin{figure}
\centerline{\psfig{figure=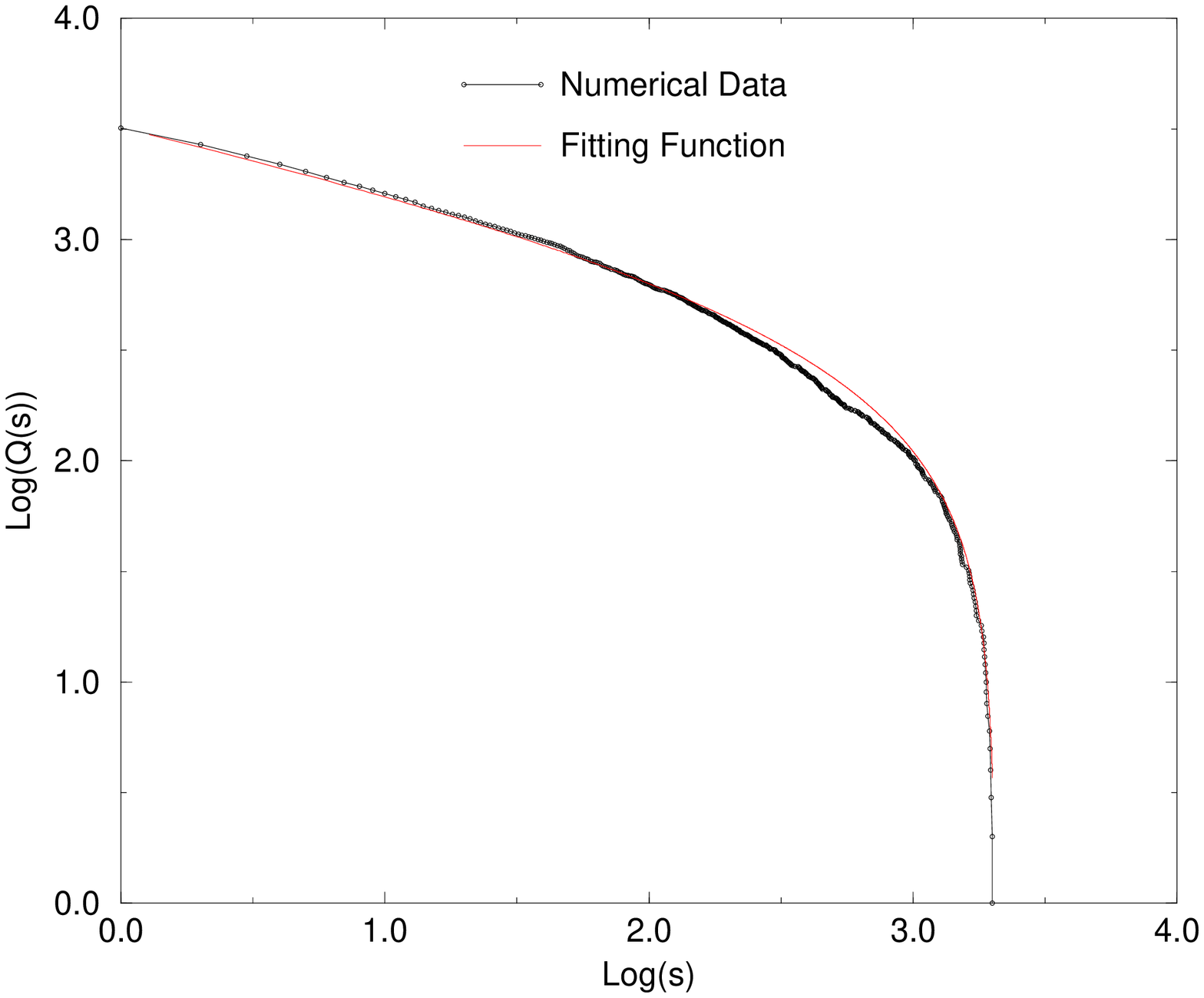,height=6.5cm,angle=270}}
\caption{$Q(s)=\int_{s}^{s_{max}} P(s')ds'$ is the integrated 
distribution; the fitting form is $As^{1-\tau'} + C$ with 
$\tau'=1.25 \pm 0.01$ ($N=1000$).}
\end{figure}
\begin{figure}
\centerline{\psfig{figure=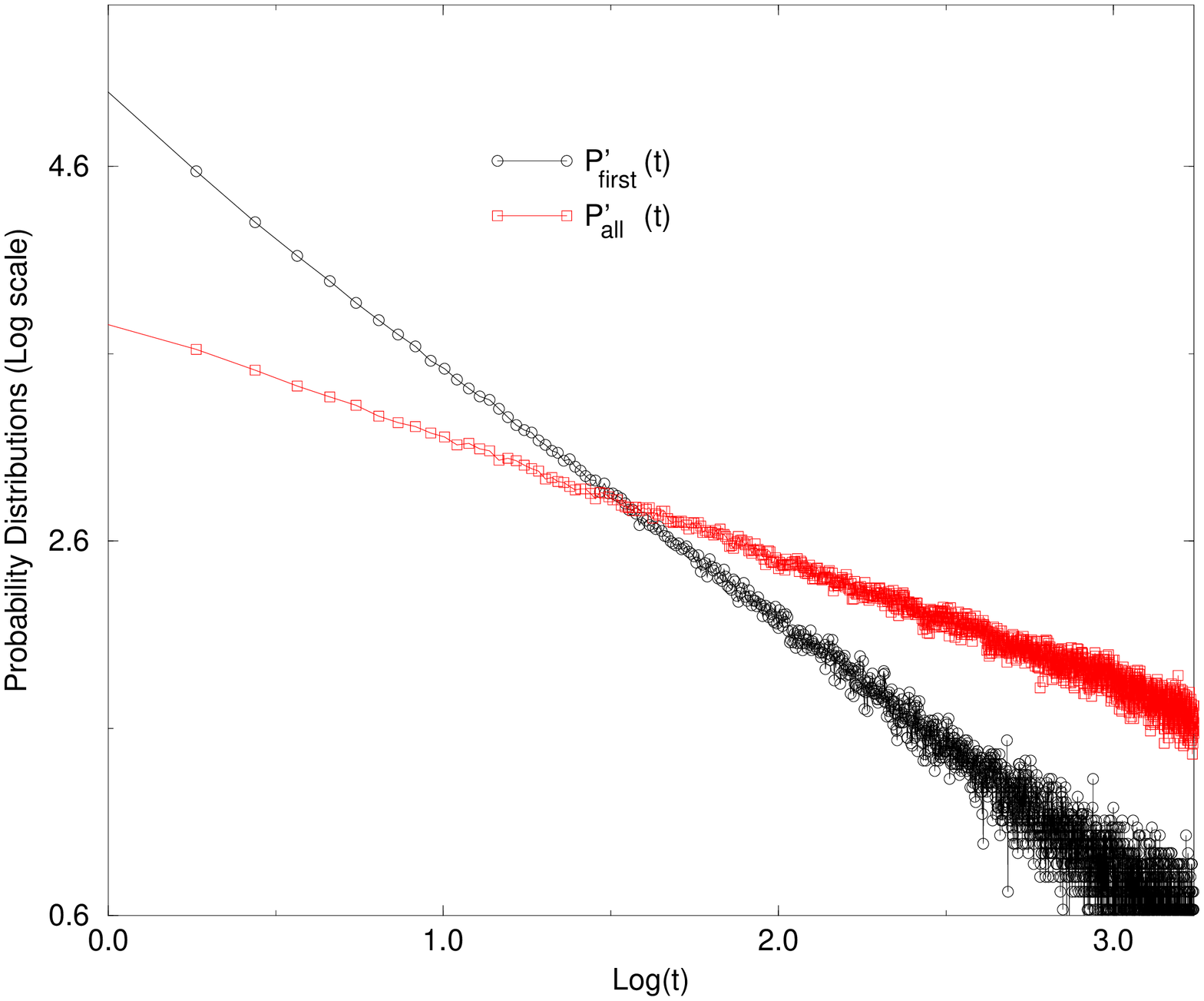,height=6.5cm,angle=270}}
\vspace{0.4cm}
\caption{{\em first} and {\em all} return times probabilities at 
boundary sites. Statistics refer to $10^9$ mutations in the whole chain.}
\end{figure}


\begin{thebibliography}{99}
\bibitem{PVZ2}
A condition analogous to our 
$G^{\prime}(1)=\sum_i ip_i=1$, is also at the basis of the RG approach to 
sandpiles of ref.[4], where it enforces energy balance.
\end{thebibliography}
\end {document}